\documentclass[a4paper,prd,showkeys,nofootinbib]{revtex4}

\usepackage[T1]{fontenc}
\usepackage[latin1]{inputenc}
\usepackage{graphicx}
\usepackage{amsmath}
\usepackage{amssymb}
\begin{document}

\title{Weyl Tensor Classification in Four-dimensional Manifolds of All Signatures}

\author{Carlos  Batista}
\email{carlosbatistas@df.ufpe.br}
\affiliation{Departamento de Física, Universidade Federal de Pernambuco, 50670-901
Recife - PE, Brazil}

\date{\today}

\begin{abstract}
It is well known that the classification of the Weyl tensor in Lorentzian manifolds of dimension four, the so called Petrov classification, was a great tool to the development of general relativity. Using the bivector approach it is shown in this article a classification for the Weyl tensor in all four-dimensional manifolds, including all signatures and the complex case, in an unified and simple way. The important Petrov classification then emerges just as a particular case in this scheme. The boost weight classification is also extended here to all signatures as well to complex manifolds. For the Weyl tensor in four dimensions it is established that this last approach produces a classification equivalent to the one generated by the bivector method.
\end{abstract}
%\thanks{CNPq financial support.}
%\archivePrefix{gr-qc/3434}

\maketitle

%\begin{center}
%   Artigo 1.2\\
%Departamento de Física da Universidade Federal de Pernambuco (DF-UFPE)
%\end{center}
%\textbf{Weyl tensor classification of four dimensional manifolds of all signatures} \\
%\\
%\\
%Author: Carlos Alberto Batista da Silva Filho
%\\
%\\
%\\
%\begin{center}
%\textbf{Abstract}\\
%\end{center}
%It is well known that the classification of Weyl tensor in Lorentzian four-dimensional manifolds, the so called %Petrov classification, was an important tool to the development of general relativity. Here it is shown that it is %possible to make a classification of Weyl tensor in a unified and simple way in all four dimensional manifolds, %including all signatures and the complex case.
%\\
%\\
%\\
%\\
%\\
%\textbf{Content}
%\begin{enumerate}
%  \item Introduction
%  \item Bivectors in 4 dimensions
%  \item Complex Weyl Tensor
%  \item Lorentzian Case, the Petrov Classification
%  \item Euclidean Signature
%  \item Signature (2,2)
%  \item Conclusion
%  \end{enumerate}

%\newpage

\section{Introduction }
Weyl tensor algebraic classification in space-times of dimension four, the Petrov classification, provides many useful techniques to deal with general relativity. In particular it can be used in the search of new solutions to Einstein's equation, the main example being \cite{typeD}, where it was found all type D vacuum solutions. This classification is also related to important geometric properties of space-times, as shows the celebrated Goldberg-Sachs theorem \cite{Goldberg-Sachs}.

There are many methods to state the Petrov classification \cite{Stephani,PlebanskiBook}, the original one was worked out by A. Z. Petrov \cite{Petrov} and is based on the fact that the Weyl tensor can be seen as an operator on the bivector space. An extensive review of this approach to the Petrov classification and a thorough analysis of the bivector space in Lorentzian four-dimensional manifolds can be found in reference \cite{HallBook}. This bivector method is the one adopted here to generalize this classification to all four-dimensional manifolds endowed with a metric and a Levi-Civita connection. The method presented in this article provides an unified local classification scheme to the Weyl tensor in four-dimensional complex manifolds and real manifolds of all signatures.

The classification of the Weyl tensor in complex manifolds of complex dimension four was already done in \cite{Plebanski75}, but using a spinorial method. Weyl tensor classification in Euclidean spaces of dimension four was also developed before \cite{Hacyan,Karlhede}, but again using different methods than the one adopted here. The (2,2) signature case was studied in \cite{Petrov-livro,Law1,Nurwoski2} using the bivector approach, but the classification schemes in these references are different from the one obtained here, while in \cite{Law2} the spinorial method was used to produce a classification that has direct relation with the one presented in this article. So the classification presented here is not new, thus the originality of the present work comes only on the approach used. It is also important to stress the unification achieved in this article, the classification of the different signatures comes easily from a common origin, this certainly helps to understand the meaning of the Weyl tensor classification and can be useful in the physical study of Wick rotated space-times. The advantages of the bivector approach are that it is simple to understand and it is useful in the analysis of the integrability of null structures on algebraically special manifolds \cite{art2}. The fact that the Weyl tensor can be seen as an operator on the bivector space is valid in all dimensions and reference \cite{ColeyBiv} took advantage of this to refine the CMPP classification of this tensor in higher-dimensional Lorentzian manifolds.

It is worth to mention that there are other forms to classify the curvature of a manifold other than the Petrov classification and its generalization presented here. The Ricci tensor, for example, can be seen as an operator on the space of vectors and the Segre classification can be used to define the different types that this operator can have \cite{Stephani}. Also the CMPP approach, which is based on the boost transformations, is a useful form to classify any tensor in Lorentzian spaces of all dimensions \cite{CMPP}. Another important method to classify the curvature is by means of the scalar invariants \cite{ColeyScInvariant}.

In section \ref{secbivectors} it will be reviewed some properties of the bivectors and of the Weyl tensor that will be required for the development of the classification. Section \ref{seccomplex} treats the complex manifolds, this case being the paradigm to the other classifications. In sections \ref{secLorentz}, \ref{secEuclid} and \ref{sec22} the classification of Lorentzian, Euclidean and (2,2) signatures are respectively done, they are seen as particular cases of the complex one. Finally, in section \ref{secboost} the classification scheme based on the boost weight obtained in \cite{ColeyPSEUD} in enhanced to include also Complex and Euclidean manifolds. It is also shown that in four dimensions the Weyl tensor classification obtained by the boost weight technique is equivalent to the one furnished by the bivector approach in all cases.
\section{Bivectors in 4 dimensions}\label{secbivectors}
Let $M$ be a differential manifold of dimension four endowed with a metric $g_{\mu\nu}$  of signature $s$. In Euclidean case $s=4$, in Lorentzian $s=2$ and in the case (2,2), where the metric can be put in the form $diag(+,+,-,-)$, the signature is $s=0$. The volume form, $\epsilon_{\mu\nu\rho\sigma}$, is a completely skew-symmetric tensor whose non-zero components in an orthonormal frame are $\pm1$. This tensor obeys the following equation:
\begin{equation}\label{epsi}
\epsilon^{\mu_1...\mu_p\nu_1...\nu_{4-p}}\, \epsilon_{\mu_1...\mu_p\sigma_1...\sigma_{4-p}} = (-1)^{\frac{4-s}{2}} \, p! (4-p)! \delta_{\sigma_1}^{[\nu_1}...\delta_{\sigma_{4-p}}^{\nu_{4-p}] }\,.
\end{equation}

A contravariant tensor of rank two which is skew-symmetric in its indices, $B^{\mu\nu} = - B^{\nu\mu} $, is called a bivector. In what follows it will also be used the term bivectors to denote the covariant and the mixed versions of these tensors. This is not inconsistent because the metric provides a map between these distinct kinds of tensors. In four dimensions the dual of a bivector $B_{\mu\nu}$ is defined by
\begin{equation}\label{dual}
    \widetilde{B}_{\mu\nu} \equiv \frac{1}{2} \epsilon_{\mu\nu\rho\sigma}B^{\rho\sigma}\,.
\end{equation}
It is easy to see that given any two bivectors, $B_{\mu\nu}$ and $F_{\mu\nu}$, we have
 \begin{equation}\label{bx}
    \widetilde{B}_{\mu\nu}F^{\mu\nu} = B_{\mu\nu}\widetilde{F}^{\mu\nu}.
\end{equation}
Taking the double dual of a bivector and using equation (\ref{epsi}) we get:
\begin{equation}\label{2dual}
    \widetilde{\widetilde{B}_{\mu\nu}} = \frac{1}{4}\epsilon_{\mu\nu\rho\sigma}\epsilon^{\rho\sigma\alpha\beta}B_{\alpha\beta} = (-1)^{\frac{s}{2}}B_{\mu\nu}\,.
\end{equation}

Let us concentrate at a specific point of the manifold, $p\in M$, and denote by $T_p$M the tangent space at this point.  From now on all considerations of this paper are at this arbitrary point. In four dimensions the space of bivectors constructed from $T_pM$  has 6 dimensions, denote this space and its complexification by $\mathfrak{B}$ and $\mathfrak{B}_{\mathbb{C}}$ respectively. The action of the dual operation can be extended form $\mathfrak{B}$ to $\mathfrak{B}_{\mathbb{C}}$ in the usual way. Let us investigate the consequences of equation (\ref{2dual}) in the various signatures and in the complex case.\\
\begin{itemize}
\item \textbf{Lorentzian signature}

In this case we shall let the duality be an operation in the complexification of the  bivector space, $\sim\,$:$\,\mathfrak{B}_{\mathbb{C}}\rightarrow\mathfrak{B}_{\mathbb{C}}$. Since $\widetilde{\widetilde{B}_{\mu\nu}} = - B_{\mu\nu} $ when $s=2$ it follows that the eigenvalues of $\sim$ are $\pm$ i. This permits to split the space $\mathfrak{B}_{\mathbb{C}}$ into a direct sum of invariant subspaces under the dual operation:
\begin{equation}\label{oplusc}
\mathfrak{B}_{\mathbb{C}} = \mathfrak{D}^+ \oplus \,\mathfrak{D}^-,
\end{equation}
$$ \mathfrak{D}^+ = \{Z^+_{\mu\nu} \in  \mathfrak{B}_{\mathbb{C}} | \widetilde{Z^+}_{\mu\nu}= i Z^+_{\mu\nu}\} \; ; \;
 \mathfrak{D}^- = \{Z^-_{\mu\nu} \in  \mathfrak{B}_{\mathbb{C}} | \widetilde{Z^-}_{\mu\nu}= -i Z^-_{\mu\nu}\}. $$
 The complex dimension of both spaces $\mathfrak{D}^+$ and $\mathfrak{D}^-$ is three. We call the former the space of self-dual bivectors and the later the space of anti-self-dual bivectors. \\
  \item \textbf{Euclidean and (2,2) signatures}

 In these cases we do not need to complexify $\mathfrak{B}$ in order to split it into the sum of invariant subspaces by the dual operation. This happens because by (\ref{2dual}) we have $\widetilde{\widetilde{B}_{\mu\nu}} =  B_{\mu\nu} $ when $s=4$ or 0, so that the eigenvalues of the duality operator, $\sim\,$:$\,\mathfrak{B}\rightarrow\mathfrak{B}$, are real, $\pm$1. This enables us to split $\mathfrak{B}$ into a direct sum of two three-dimensional invariant subspaces under the dual operation:
$$\mathfrak{B} = \mathfrak{D}^+ \oplus \,\mathfrak{D}^-,$$
$$ \mathfrak{D}^+ = \{Z^+_{\mu\nu} \in  \mathfrak{B} | \widetilde{Z^+}_{\mu\nu}=  Z^+_{\mu\nu}\} \; ; \;
 \mathfrak{D}^- = \{Z^-_{\mu\nu} \in  \mathfrak{B}| \widetilde{Z^-}_{\mu\nu}= - Z^-_{\mu\nu}\}. $$

 \item \textbf{Complex case}\\
 In complex manifolds of complex dimension four the duality operator is such that its square can give both results, the identity or minus the identity. This happens because in this kind of manifold vectors of an orthonormal frame can be multiplied by a factor of $i$, changing the apparent signature of the metric. The important thing is that once a choice of volume form is made, the bivector space $\mathfrak{B}_{\mathbb{C}}$, can be split in a direct sum of invariant subspaces, under the duality operation, of complex dimension three, as in equation (\ref{oplusc}). Here it will be assumed, without loss of generality, that the volume form is conveniently chosen is such a way that the dual operation squared gives the identity map. \\
\end{itemize}
 Note that in the various signatures the same symbols, $\mathfrak{D}^+$ and $\mathfrak{D}^-$, were used to denote different spaces. This shall make no confusion since it will be clear in the context which space it is meant. It is worth to keep in mind that there is nothing intrinsic to the manifold which distinguishes these two spaces, it is just a choice of orientation sign. If we change the sign of $\epsilon_{\mu\nu\rho\sigma}$ the spaces $\mathfrak{D}^+$ and $\mathfrak{D}^-$ are interchanged. Note also that (\ref{bx}) implies that if $Z^+_{\mu\nu}\in\mathfrak{D}^+$ and  $Z^-_{\mu\nu}\in\mathfrak{D}^-$ then $Z^+_{\mu\nu}Z^{-\mu\nu} = 0 $. Until the end of this section the calculations are valid to all signatures and also to the complex case.
\\

  The Weyl tensor of the manifold $(M,g_{\mu\nu})$ has the following symmetries\footnote{It is assumed that the connection on the manifold is torsion-free and compatible with the metric, the so called Levi-Civita connection.}:
 \begin{equation}\label{weyl}
 C_{\mu\nu\rho\sigma} = C_{[\mu\nu]\rho\sigma} = C_{\mu\nu[\rho\sigma]} = C_{\rho\sigma\mu\nu}\; ; \;
 C^\mu_{\phantom{\mu}\nu\mu\sigma} = 0 \; ; \; C_{\mu[\nu\rho\sigma]} = 0.
 \end{equation}
 Because of the skew-symmetry in the first and second pairs of the Weyl tensor indices it is natural to define
 \begin{equation}\label{Weyldualdefinition}
    C_{\mu\nu\widetilde{\rho\sigma}} \equiv \frac{1}{2} \epsilon_{\rho\sigma\alpha\beta} C_{\mu\nu}^{\phantom{\mu\nu}\alpha\beta}\; ; \; C_{\widetilde{\mu\nu}\rho\sigma} \equiv \frac{1}{2} \epsilon_{\mu\nu\alpha\beta} C^{\alpha\beta}_{\phantom{\alpha\beta}\rho\sigma}.
 \end{equation}
 It is then trivial to see that $C_{\mu\nu\widetilde{\rho\sigma}} = C_{\widetilde{\rho\sigma}\mu\nu}$, but it less obvious that $C_{\mu\nu\widetilde{\rho\sigma}} = C_{\widetilde{\mu\nu}\rho\sigma}$, let us prove this.
 $$
 \textrm{Define}, \; \; T_{\mu\nu\rho\sigma} \equiv C_{\mu\nu\widetilde{\rho\sigma}} - C_{\widetilde{\mu\nu}\rho\sigma}.
  $$
  Note that $T_{\mu\nu\rho\sigma} = T_{[\mu\nu]\rho\sigma} = T_{\mu\nu[\rho\sigma]} =- T_{\rho\sigma\mu\nu}$.
  $$\epsilon^{\mu\rho\alpha\beta}T_{\mu\nu\rho\sigma} = \epsilon^{\mu\rho\alpha\beta}[\frac{1}{2}\epsilon_{\rho\sigma\theta\gamma}C_{\mu\nu}^{\phantom{\mu\nu}\theta\gamma} - \frac{1}{2}\epsilon_{\mu\nu\theta\gamma}C^{\theta\gamma}_{\phantom{\theta\gamma}\rho\sigma}] =
  $$
   \begin{equation}\label{epsiT}
   = -3(-1)^{\frac{s}{2}}(\delta_{\sigma}^{[\mu}\delta_\theta^\alpha\delta_\gamma^{\beta]}C_{\mu\nu}^{\phantom{\mu\nu}\theta\gamma} + \delta_{\nu}^{[\rho}\delta_\theta^\alpha\delta_\gamma^{\beta]}C^{\theta\gamma}_{\phantom{\theta\gamma}\rho\sigma}) = -(-1)^{\frac{s}{2}}(C_{\sigma\nu}^{\phantom{{\sigma\nu}}\alpha\beta}+C^{\alpha\beta}_{\phantom{\alpha\beta}\nu\sigma}) = 0\,,
   \end{equation}
 where in the second equality it was used (\ref{epsi}) and in the third (\ref{weyl}).  In particular, equation (\ref{epsiT}) means that
  $T_{\mu\nu\rho\sigma} =  T_{\rho\nu\mu\sigma}$. Further if we contract $\alpha$ and $\nu$ in equation (\ref{epsiT}) we get $T_{[\mu\nu\rho]\sigma} = 0$. Using the symmetries of $T_{\mu\nu\rho\sigma}$ in this equation it is easily seen that this tensor vanishes, which implies the wanted relation. Summing-up we have,
  \begin{equation}\label{weyld}
    C_{\mu\nu\widetilde{\rho\sigma}} = C_{\widetilde{\rho\sigma}\mu\nu} \;\; ; \; \; C_{\mu\nu\widetilde{\rho\sigma}} = C_{\widetilde{\mu\nu}\rho\sigma}.
  \end{equation}
  The above identities are the most important relations of this paper, since from them it is trivial to deduce that the Weyl operator, to be defined next section, has the interesting property of sending elements of $\mathfrak{D}^{\pm}$ into elements of $\mathfrak{D}^{\pm}$.
  \section{Complex Weyl Tensor}\label{seccomplex}
The intent of this section is to produce an algebraic classification for the Weyl tensor in complex manifolds. This was already done in \cite{Plebanski75} by means of Penrose's spinor techniques, while here the bivector approach is taken, but at the end the possible algebraic types are the same. The complex case will be used in forthcoming sections to  extract the Weyl tensor classification in real manifolds of arbitrary signature in an easy and quick way.

     Let us assume that $(M,g_{\mu\nu})$ is a complex manifold of complex dimension four. The metric can be complex, so that the Weyl tensor is, in general, not real. The classification scheme done in this article is based on the possibility of seeing the Weyl tensor as an operator on the complex bivector space, $C:\, \mathfrak{B}_{\mathbb{C}} \rightarrow\mathfrak{B}_{\mathbb{C}}$, with action given by:
  \begin{equation}\label{Weyl operator}
   C(B)=F\;\;\textrm{where}\;\; F_{\mu\nu} \equiv C_{\mu\nu\rho\sigma}B^{\rho\sigma}.
  \end{equation}
  By making an analysis of the non-trivial solutions of the eigenvalue equation $C(B)= \lambda B$ it is possible to classify the Weyl tensor. But this approach may be laborious if the Weyl operator action is studied in the whole $\mathfrak{B}_{\mathbb{C}}$, as A. Z. Petrov did in the Lorentzian case \cite{Petrov}. However a shortcut can be taken using the restrictions of this operator to the subspaces of self-dual and anti-self-dual bivectors. This makes possible to attack this eigenvalue problem working with two $3\times3$ matrices instead of the $6\times6$ matrix which represents the Weyl operator on the full bivector space.

  Many calculations in this section will be done in such a way to include not only the complex manifold case, but also real manifolds of all signatures, this shall be clear in the context. Let $Z^+_{\mu\nu}$ be a self-dual bivector, this means that $\widetilde{Z^+}_{\mu\nu} = \varepsilon Z^+_{\mu\nu}$, where $\varepsilon$ is $+i$ in Lorentzian signature and $+1$ in the other two signatures as well in the complex case. Then if $T_{\mu\nu}$ is the image of this bivector under the Weyl operator, $T_{\mu\nu}=C_{\mu\nu\rho\sigma}Z^{+\rho\sigma}$, it must be self-dual:
  \begin{equation}\label{Weyl Z into Z}
    \widetilde{T}_{\mu\nu} = C_{\widetilde{\mu\nu}\rho\sigma}Z^{+\rho\sigma} =C_{\mu\nu\widetilde{\rho\sigma}}Z^{+\rho\sigma} = C_{\mu\nu\rho\sigma}\widetilde{Z}^{+\rho\sigma} = \varepsilon C_{\mu\nu\rho\sigma}Z^{+\rho\sigma} \equiv \varepsilon T_{\mu\nu}.
  \end{equation}
  Where in the second equality it was used equation (\ref{weyld}) , in the third eq. (\ref{bx}) and in the fourth it was used the self-duality of $Z^+_{\mu\nu}$. Equation (\ref{Weyl Z into Z}) implies that $T_{\mu\nu}$ is self-dual. In an analogous way this operator sends anti-self-dual bivectors into anti-self-dual bivectors. This together with (\ref{oplusc}) guarantees that the Weyl operator is the direct sum of its restrictions to $\mathfrak{D}^+$ and $\mathfrak{D}^-$,
  \begin{equation}\label{C partition}
    C_{\mu\nu\rho\sigma} = C^+_{\mu\nu\rho\sigma} + C^-_{\mu\nu\rho\sigma}\; \; ; \;C^{\pm}_{\mu\nu\rho\sigma} = \frac{1}{2}(C_{\mu\nu\rho\sigma} \pm \varepsilon^3 C_{\mu\nu\widetilde{\rho\sigma}}).
  \end{equation}
Where $C^+(C^-) $ is called the self-dual(anti-self-dual) part of the Weyl tensor, its action in $\mathfrak{D}^-$ ($\mathfrak{D}^+$) gives zero. For example, if $Z^-_{\mu\nu}$ is an element of $\mathfrak{D}^-$ then using (\ref{bx}) and the relation $\widetilde{Z^-}=-\varepsilon Z^-$, we have:
\begin{equation*}
    C^+_{\mu\nu\rho\sigma}Z^{-\rho\sigma} =  \frac{1}{2}(C_{\mu\nu\rho\sigma}Z^{-\rho\sigma} + \varepsilon^3 C_{\mu\nu\rho\sigma}\widetilde{Z}^{-\rho\sigma}) = \frac{1}{2}(C_{\mu\nu\rho\sigma} + \varepsilon^3 (-\varepsilon) C_{\mu\nu\rho\sigma})Z^{-\rho\sigma} = 0\,.
\end{equation*}
Thus to classify the Weyl tensor is equivalent to classify the operators $C^+:\mathfrak{D}^+\rightarrow\mathfrak{D}^+$ and  $C^-:\mathfrak{D}^-\rightarrow\mathfrak{D}^-$. Now let us see that the these two operators have vanishing trace. Using equations (\ref{weyl}), (\ref{Weyldualdefinition}) and (\ref{C partition}) we see that:
$$C^{\pm \mu}_{\phantom{\pm \mu}\nu\mu\sigma} = \frac{1}{2}(C^\mu_{\phantom{\mu}\nu\mu\sigma} \pm \varepsilon^3 \frac{1}{2}\epsilon_{\mu\sigma\alpha\beta}C^{\mu\phantom{\nu}\alpha\beta}_{\phantom{\mu}\nu}) = \mp \varepsilon^3 \frac{1}{4}\epsilon_{\mu\sigma\alpha\beta}C_\nu^{\phantom{\nu}[\mu\alpha\beta]} = 0.$$
In particular this implies that $C^{\pm \mu\nu}_{\phantom{\pm \mu\nu}\mu\nu} =0$, proving that the operators $C^\pm$ have zero trace. Below it will be explicitly shown that in complex manifolds these two operators are independent of each other and have in general no simplifying property other than the vanishing trace. Thus the Weyl tensor classification in complex manifolds reduces to the classification of two trace-free operators acting on three-dimensional spaces. The three eigenvalues, $(\lambda_1,\lambda_2,\lambda_3)$, of a trace-free operator acting on a three-dimensional vector space can have the following features: (a) All eigenvalues are different, $(\lambda_1,\lambda_2,-\lambda_1-\lambda_2)$; (b) There is a non-zero repeated eigenvalue, $(\lambda,\lambda,-2\lambda)$ (c) All eigenvalues are zero, $(0,0,0)$. In the case (a) the operator is called type I, in the case (b) there are two algebraically distinct possibilities, if the operator can be put in a diagonal form it is type D, otherwise it is type II, finally the case (c) allows three distinctions, when the operator vanishes it is called type O, when it is non-zero but with vanishing square it is type N and when its square is different from zero it is said to be type III. These types are summarized by the bellow algebraic characteristics:\\ \\
  $\left\{
  \begin{array}{ll}
    \textbf{Type  O}^{\pm} \rightarrow \;  C^\pm = 0  \\
    \textbf{Type  I}^{\pm} \;\,\rightarrow \; C^\pm $ allows 3 distinct eigenvalues  $  \\
    \textbf{Type  D}^{\pm} \rightarrow  C^\pm $ is diagonalizable with a repeated non-zero eigenvalue   $ \\
    \textbf{Type  II}^{\pm} \rightarrow  C^\pm $ is non-diagonalizable with a repeated non-zero eigenvalue$ \\
   \textbf{Type  III}^{\pm} \rightarrow   \, (C^\pm)^3 = 0\, $and$ \,(C^+)^2 \neq 0 $\,(all eigenvalues are zero)$ \\
    \textbf{Type  N}^{\pm} \rightarrow   \, (C^\pm)^2 = 0 \,$and$\, C^+ \neq 0 $ \,(all eigenvalues are zero)$
  \end{array}
\right.$\\ \\ \\
The classification of the full Weyl tensor is then made by a composition of these algebraic types. For example, we say that this tensor is type (I,O) if $C^+$ is type I$^+$ and $C^-$ is type O$^-$. But it is important to note that type (I,O) is intrinsically equivalent to type (O,I), since the operators $C^+$ and $C^-$ are interchanged by a simple change of orientation sign, and from an intrinsic point of view this sign choice is arbitrary. So at the end there are 21 possible classifications:
\begin{center}
$\textrm{(O,O); (O,I); (O,D);(O,II); (O,III); (O,N); (I;I); (I;D); (I,II); (I,III); (I,N);}$\\
$\textrm{(D;D); (D,II); (D,III); (D,N); (II,II); (II,III); (II,N); (III,III); (III,N); (N,N).}$\\
\end{center}
It is worth to note that such classification was made at a specific point of the manifold, $p\in M$, and in general the Weyl tensor can change its type from point to point. Now to understand better this classification let us find explicit representations for the operators $C^\pm$. In the complex tangent space it is always possible to construct a null tetrad frame, $\{l,n,m_1,m_2\}$, defined to be such that the only non-zero contractions are
\begin{equation}\label{NulltetradContrac}
    l^\mu n_\mu = 1 \;\; \textrm{and} \;\; m_1^\mu m_{2\mu} = -1\,.
\end{equation}
In particular all basis vectors are null. For example, if $\{e_{(1)},e_{(2)},e_{(3)},e_{(4)}\}$ is an orthonormal frame such that $e_{(a)}^\mu e_{(b)\mu}=\delta_{ab}$, then the vectors $l=\frac{1}{\sqrt{2}}(e_{(1)}+ie_{(2)}) \, , \, n=\frac{1}{\sqrt{2}}(e_{(1)}-ie_{(2)})\,,\, m_1=\frac{1}{\sqrt{2}}(e_{(3)}+ie_{(4)}) \;\textrm{and}\;m_2=\frac{-1}{\sqrt{2}}(e_{(3)}-ie_{(4)})\,$ form a null tetrad frame.

The Weyl tensor in a real four-dimensional manifold has 10 independent real components, because of the symmetries shown in equation (\ref{weyl}), while in complex manifolds it has 10 independent complex components. In the complex case it is convenient to choose these independent components to be

$$ \Psi^+_0 \equiv C_{\mu\nu\rho\sigma}l^\mu m_1^\nu l^\rho m_1^\sigma \; ; \; \Psi^+_1 \equiv C_{\mu\nu\rho\sigma}l^\mu n^\nu l^\rho m_1^\sigma \; ; \; \Psi^+_2 \equiv C_{\mu\nu\rho\sigma}l^\mu m_1^\nu m_2^\rho n^\sigma  $$
\begin{equation}\label{WeylScalars}
\Psi^+_3 \equiv C_{\mu\nu\rho\sigma}l^\mu n^\nu m_2^\rho n^\sigma \; ; \;\Psi^+_4 \equiv C_{\mu\nu\rho\sigma}n^\mu m_2^\nu n^\rho m_2^\sigma
\end{equation}
$$ \Psi^-_0 \equiv C_{\mu\nu\rho\sigma}l^\mu m_2^\nu l^\rho m_2^\sigma \; ; \; \Psi_1^- \equiv C_{\mu\nu\rho\sigma}l^\mu n^\nu l^\rho m_2^\sigma \; ; \; \Psi_2^- \equiv C_{\mu\nu\rho\sigma}l^\mu m_2^\nu m_1^\rho n^\sigma  $$
$$\Psi_3^- \equiv C_{\mu\nu\rho\sigma}l^\mu n^\nu m_1^\rho n^\sigma \; ; \;\Psi_4^- \equiv C_{\mu\nu\rho\sigma}n^\mu m_1^\nu n^\rho m_1^\sigma\,.   $$
These scalars are called the Weyl scalars. The main advantage of using a null tetrad frame when dealing with the Weyl tensor is that the trace-free property and the Bianchi identity, (\ref{weyl}), are easily written. Equations below are the explicit forms of these two properties of the Weyl tensor in this kind of frame:  %They are summarized in the below equations:
\begin{eqnarray}\label{WeylSimNulltetrad}
% \nonumber to remove numbering (before each equation)
  \nonumber C_{1l1n} &=& C_{2l2n}= C_{l1l2}= C_{n1n2} = 0 \; ; \\
  C_{1l12}  &=& \Psi^+_1\;;\; C_{2l21} = \Psi_1^-\;;\;C_{1n12} = \Psi_3^-\;;\;C_{2n21} = \Psi^+_3\;; \\
  \nonumber C_{1212}&=& C_{lnln} =  \Psi^+_2+\Psi_2^-\; ; \; C_{12ln} = \Psi_2^--\Psi^+_2.
\end{eqnarray}

Where, for example, $C_{12ln}$ means $C_{\mu\nu\rho\sigma}m^\mu_1m^\nu_2l^\rho n^\sigma$. In particular, from equation (\ref{WeylSimNulltetrad}) we see that the Weyl scalars, defined in (\ref{WeylScalars}), are indeed independent of each other in complex manifolds, so that they can be used to represent the Weyl tensor degrees of freedom. Choosing the manifold  orientation to be such that $\epsilon_{\mu\nu\rho\sigma}e^\mu_{(1)}e^\nu_{(2)}e^\rho_{(3)}e^\sigma_{(4)}=1,$ then convenient bases for spaces $\mathfrak{D}^+$ and $\mathfrak{D}^-$ are respectively given by:
\begin{equation}\label{Zdual}
    Z^{+1}_{\mu\nu} = 2l_{[\mu}m_{1\nu]} \; ; \; Z^{+2}_{\mu\nu} = 2m_{2[\mu}n_{\nu]} \; ; \; Z^{+3}_{\mu\nu} = \sqrt{2} n_{[\mu}l_{\nu]} + \sqrt{2} m_{1[\mu}m_{2\nu]}
\end{equation}
\begin{equation}\label{Xantidual}
    Z^{-1}_{\mu\nu} = 2l_{[\mu}m_{2\nu]} \; ; \; Z^{-2}_{\mu\nu} = 2m_{1[\mu}n_{\nu]} \; ; \; Z^{-3}_{\mu\nu} = \sqrt{2} n_{[\mu}l_{\nu]} + \sqrt{2} m_{2[\mu}m_{1\nu]}.
\end{equation}
    The only non-zero full contractions of the elements of these bases are:
    \begin{equation}\label{ZcontracXcontr}
       Z^{+1\mu\nu}Z^{+2}_{\mu\nu} = 2 \; ; \;  Z^{+3\mu\nu}Z^{+3}_{\mu\nu} = -2 \; ; \;Z^{-1\mu\nu}Z^{-2}_{\mu\nu} = 2 \;; \; Z^{-3\mu\nu}Z^{-3}_{\mu\nu} = -2\,.
    \end{equation}
 Using equations (\ref{WeylScalars}), (\ref{WeylSimNulltetrad}), (\ref{Zdual}) and (\ref{Xantidual}) it is straightforward to see that the Weyl scalars can be written in the form
 \begin{eqnarray}\label{contractions}
                          \nonumber
                           \Psi^\pm_0 &=& \frac{1}{4}C_{\mu\nu\rho\sigma}Z^{\pm1\mu\nu}Z^{\pm1\rho\sigma}    \;  ; \; \Psi^\pm_1 = \frac{-\sqrt{2}}{8}C_{\mu\nu\rho\sigma}Z^{\pm1\mu\nu}Z^{\pm3\rho\sigma} \\
                            \Psi^\pm_2 &=& \frac{1}{4}C_{\mu\nu\rho\sigma}Z^{\pm1\mu\nu}Z^{\pm2\rho\sigma} = \frac{1}{8}C_{\mu\nu\rho\sigma}Z^{\pm3\mu\nu}Z^{\pm3\rho\sigma} \\  \nonumber
                           \Psi^\pm_3 &=& \frac{-\sqrt{2}}{8}C_{\mu\nu\rho\sigma}Z^{\pm2\mu\nu}Z^{\pm3\rho\sigma}    \;  ; \; \Psi^\pm_4 = \frac{1}{4}C_{\mu\nu\rho\sigma}Z^{\pm2\mu\nu}Z^{\pm2\rho\sigma}.
                         \end{eqnarray}
   The matrix representations of operators $C^\pm$ in these bivector bases are $ \mathcal{C}^{\pm\phantom{j}i}_{\;\;j}$, defined to be such that $ C^\pm (Z^{\pm i}) = \mathcal{C}^{\pm\phantom{j}i}_{\;\;j}Z^{\pm j}$ or, in component form, $\mathcal{C}^{\pm\phantom{j}i}_{\;\;j}Z^{\pm j}_{\mu\nu} = C_{\mu\nu\rho\sigma}Z^{\pm i\,\rho\sigma}$. Where it was used the fact that $C^\pm_{\mu\nu\rho\sigma}Z^{\pm i\,\rho\sigma} =C_{\mu\nu\rho\sigma}Z^{\pm i\,\rho\sigma}$, which stems from (\ref{C partition}) and from the triviality of $C^\pm$ action on the spaces of opposite duality. Now using these definitions and equations (\ref{ZcontracXcontr}) and (\ref{contractions}) we get that the matrix representations of operators $C^\pm$ are:
\begin{equation}\label{Cmatrix}
\mathcal{C}^{\pm\phantom{j}i}_{\;\;j} = 2\left[
                                                                             \begin{array}{ccc}
                                                                               \Psi^\pm_2 & \Psi^\pm_4 & -\sqrt{2}\Psi^\pm_3 \\
                                                                               \Psi^\pm_0& \Psi^\pm_2 & -\sqrt{2}\Psi^\pm_1 \\
                                                                               \sqrt{2}\Psi^\pm_1 & \sqrt{2}\Psi^\pm_3 & -2\Psi^\pm_2 \\
                                                                             \end{array}
                                                                           \right].
\end{equation}
Note that these matrices have vanishing trace, as it should be. In next section these matrix representations will be used to find a canonical form for each classification type seen above. Complex results will be used to generate a Weyl tensor classification on real four-dimensional manifolds in the forthcoming sections.
\section{Lorentzian Case, The Petrov Classification}\label{secLorentz}
Real Lorentzian manifolds are characterized by the local existence of a real frame \{$e_t,e_x,e_y,e_z$\} such that the only non-zero contractions are $e^\mu_t e_{t\mu} = 1$ and $e^\mu_x e_{x\mu} = e^\mu_y e_{y\mu} = e^\mu_z e_{z\mu} =-1$. This kind of frame is called a Lorentz frame or a tetrad. From these real vectors it is possible to construct a null tetrad frame in the complexified tangent space, $\mathbb{C}\otimes T_pM$, given by $l=\frac{1}{\sqrt{2}}(e_t+e_z) \, , \, n=\frac{1}{\sqrt{2}}(e_t-e_z)\, , \, m_1=\frac{1}{\sqrt{2}}(e_x+ie_y) \; \textrm{and} \;m_2=\frac{1}{\sqrt{2}}(e_x-ie_y)$.

Since this null tetrad frame has exactly the same inner products as the null tetrad frame seen in the previous section it follows that basically the whole formalism is the same in the present case. The differences that must be stressed are that now  the Weyl tensor is assumed to be real, $\overline{C}_{\mu\nu\rho\sigma}=C_{\mu\nu\rho\sigma}$, and the basis vectors $l$ and $n$ are real while $m_1$ and $m_2$ are complex and conjugates to each other,  $\overline{m_1} = m_2$. These simple observations make huge restrictions on the possible classifications of the Weyl tensor, since now the Weyl scalars $\Psi^+_a$ and $\Psi^-_a$ are complex conjugates to each other, $\overline{\Psi^+_a} = \Psi^-_a$, which stems from their definitions, (\ref{WeylScalars}). This reduces the number of independent components of the Weyl tensor from 10 complex numbers to 5 complex numbers, or 10 real parameters, as it should be in a real manifold.

So by equation (\ref{Cmatrix}) we have that $\mathcal{C}^{+\phantom{j}i}_{\;j}$ and $\mathcal{C}^{-\phantom{j}i}_{\;j}$ are complex conjugates to each other\footnote{In particular this implies that if $\lambda$ is an eigenvalue of the full Weyl operator then its complex conjugate, $\overline{\lambda}$, is also an eigenvalue. This result was proved by A. Z. Petrov in \cite{Petrov}. }. This implies that if the operator $C^+$ has certain algebraic type then the operator $C^-$ will have the same type. Thus the only allowable classifications are (O,O); (I,I);(D,D); (II,II); (III,III) and (N,N). The notation can now be condensed and we say that the Weyl tensor in a real Lorentzian manifold can have just six types of classification: O, I, D, II, III and N, where type D, for example, means (D,D). These types are the well known Petrov types, which are widely studied since its discovery in 1954 \cite{Petrov}\footnote{When created by A. Z. Petrov the classification consisted only of three non-trivial types, called types 1, 2 and 3. Type 1 later was split into types I and D, type 2 was refined to give types II and N, while type 3 was the now called type III \cite{Bel}.}.

There are various ways to approach the Petrov classification \cite{Stephani}, probably the most elegant is Penrose's spinor method \cite{Penrose}. It is worth to make a connection between the classification as presented in this text and some of the other ones. In spinorial approach it is easy to see that the different Petrov types are featured by the possibility of choosing convenient bases where some Weyl scalars are zero. The final result of this analysis is summarized in the table below.
\begin{center}
\textmd{Weyl Scalars that Can be Made to Vanish by a Suitable Choice of Basis}\\
\begin{tabular}{|c|c|c|}

  \hline
  %after \\: \hline or \cline{col1-col2} \cline{col3-col4} ...

  Type O - All & Type I - $\Psi^\pm_0,\Psi^\pm_4$ & Type D - $\Psi^\pm_0,\Psi^\pm_1,\Psi^\pm_3,\Psi^\pm_4$  \\ \hline
  Type II -$\Psi^\pm_0,\Psi^\pm_1,\Psi^\pm_4$ & Type III - $\Psi^\pm_0,\Psi^\pm_1,\Psi^\pm_2,\Psi^\pm_4 \>$ & Type N - $\Psi^\pm_0,\Psi^\pm_1,\Psi^\pm_2,\Psi^\pm_3$  \\
  \hline
\end{tabular}
\end{center}

This approach to the characterization of Petrov types, which identify the many types with the possibility to annihilate different Weyl tensor components, is the origin of the main extension of this classification to higher dimensions, the so called CMPP classification \cite{CMPP}. Now it is easy matter to see that the classification obtained here using the algebraic properties of the operators $C^\pm$ is equivalent to the classification depicted in the above table. For example, if the Weyl tensor is Petrov type N according to the above table it is possible, using equation (\ref{Cmatrix}), to find a basis such that the representations of $C^+$ and $C^-$ are respectively:
\begin{equation*}
    \mathcal{C}^{+\phantom{j}i}_{\;\;j} = 2\left[
                                                                             \begin{array}{ccc}
                                                                               0 & \Psi^+_4 & 0\\
                                                                               0& 0 & 0 \\
                                                                               0 & 0 & 0 \\
                                                                             \end{array}
                                                                           \right] \; \;\textrm{and} \; \; \mathcal{C}^{-\phantom{j}i}_{\;\;j} = 2\left[
                                                                             \begin{array}{ccc}
                                                                               0 & \overline{\Psi^+_4} & 0\\
                                                                               0& 0 & 0 \\
                                                                               0 & 0 & 0 \\
                                                                             \end{array}
                                                                           \right].
\end{equation*}
Where $\Psi^+_4 \neq 0$ and it was used the fact that $\overline{\Psi^+_a} = \Psi^-_a$ in Lorentzian signature. It is now trivial to see that $(C^+)^{2} = 0 =(C^-)^{2}$, which means that the Weyl tensor, according to classification defined in section \ref{seccomplex}, is type (N,N), or type N in the condensed notation. This illustrates the equality of type N Petrov classification in both approaches. It is also easy to verify the equivalence of the other types.

 As a final comment, note that the above table enables to determine the canonical forms of operators $C^\pm$ matrix representations in the various classification types. For example, type (II,II) is characterized by the following canonical forms:
\small{ \begin{equation*}
 \mathcal{C}^{+\phantom{j}i}_{\;\;j} = 2\left[
                                                                             \begin{array}{ccc}
                                                                               \Psi^+_2 & 0 & -\sqrt{2}\Psi^+_3\\
                                                                               0& \Psi^+_2 & 0 \\
                                                                               0 & \sqrt{2}\Psi^+_3 & -2\Psi^+_2 \\
                                                                             \end{array}
                                                                           \right] \; \;\textrm{and} \; \; \mathcal{C}^{-\phantom{j}i}_{\;\;j} = 2\left[
                                                                             \begin{array}{ccc}
                                                                               \overline{\Psi^+_2} & 0 & -\sqrt{2}\overline{\Psi^+_3}\\
                                                                               0& \overline{\Psi^+_2} & 0 \\
                                                                               0 & \sqrt{2}\overline{\Psi^+_3} & -2\overline{\Psi^+_2} \\
                                                                             \end{array}
                                                                           \right],
\end{equation*}}\normalsize{with $\Psi^+_2\neq0\neq\Psi^+_3$.} Since a complex manifold of complex dimension four can always be seen locally as a complexified Lorentzian manifold of real dimension four, it follows that canonical forms to the complex case can also be easily found by using the above table. For example, if the Weyl tensor on a complex manifold has type (D,III) then it can be found a basis for space $\mathfrak{D}^+$ and an independent basis  for $\mathfrak{D}^-$ (the bases of these two spaces may be not related to each other as they are in equations (\ref{Zdual}) and (\ref{Xantidual})) where the representations of the operators $C^\pm$ are:
\small{ \begin{equation*}
 \mathcal{C}^{+\phantom{j}i}_{\;\;j} = 2\left[
                                                                             \begin{array}{ccc}
                                                                               \Psi^+_2 & 0 & 0\\
                                                                               0& \Psi^+_2 & 0 \\
                                                                               0 & 0 & -2\Psi^+_2 \\
                                                                             \end{array}
                                                                           \right] \; \;\textrm{and} \; \; \mathcal{C}^{-\phantom{j}i}_{\;\;j} = 2\left[
                                                                             \begin{array}{ccc}
                                                                               0 & 0 & -\sqrt{2}\Psi^-_3\\
                                                                               0& 0 & 0 \\
                                                                               0 & \sqrt{2}\Psi^-_3 & 0 \\
                                                                             \end{array}
                                                                           \right].
\end{equation*}}\normalsize{As a last comment note also that by definition of types I$^\pm$ in the preceding section, the matrix representations of the operators $C^\pm$ with these types can always be put in a diagonal form with three different eigenvalues whose sum is zero. So $diag(\lambda^\pm_1,\lambda^\pm_2,\lambda^\pm_3)$, with $\lambda^\pm_i\neq\lambda^\pm_j$ if $i\neq j$ and $\lambda^\pm_1+\lambda^\pm_2+\lambda^\pm_3=0$, may be convenient choices of canonical forms for the types I$^\pm$.
\section{Euclidean Signature}\label{secEuclid}
Now suppose that $(M,g_{\mu\nu})$  is an Euclidean} four-dimensional real manifold. In this case it is possible to find an orthonormal frame $\{e_{(1)},e_{(2)},e_{(3)},e_{(4)}\}$ made of real vectors such that  $e_{(a)}^\mu e_{(b)\mu}=\delta_{ab}$. Thus a null tetrad frame, $\{l,n,m_1,m_2\}$, can be constructed in the complexified tangent space by defining the vectors $l=\frac{1}{\sqrt{2}}(e_{(1)}+ie_{(2)}) \, , \, n=\frac{1}{\sqrt{2}}(e_{(1)}-ie_{(2)})\,,\, m_1=\frac{1}{\sqrt{2}}(e_{(3)}+ie_{(4)}) \; \textrm{and} \;m_2=\frac{-1}{\sqrt{2}}(e_{(3)}-ie_{(4)})\,$.

Since $\{e_{(a)},\small{a=1,2,3,4}\}$ \normalsize{ are real vectors it follows that $l$ is the complex} conjugate of $n$, while $m_1$ is minus the complex conjugate of $m_2$. Also the Weyl tensor is real, because it is being assumed that the metric is real. Using these relations it is easy to see that the Weyl scalars, (\ref{WeylScalars}), are such that: $\overline{\Psi^{\pm}_0} = \Psi^{\pm}_4$ , $\overline{\Psi^{\pm}_1} = -\Psi^{\pm}_3$ and $\overline{\Psi^{\pm}_2} = \Psi^{\pm}_2$. These equalities together with equation (\ref{Cmatrix}) implies that the matrix representations of $C^+$ and $C^-$ are hermitian and independent of each other. Since these matrices have vanishing trace and every hermitian matrix can be put in a diagonal form by a suitable choice of basis, it follows that there are just three algebraically distinct canonical forms for the matrix representations of $C^+$ and $C^-$, they are: $diag(\lambda^\pm_1,\lambda^\pm_2,-\lambda^\pm_1-\lambda^\pm_2)$ with all eigenvalues different,  $diag(\lambda^\pm,\lambda^\pm,-2\lambda^\pm)$ with $\lambda^\pm\neq0$ and $diag(0,0,0)$. This means that $C^\pm$ can be just of types I$^\pm$, D$^\pm$ or O$^\pm$. Since the operators $C^+$ and $C^-$ are independent of each other, there are also just six realizable classification types in the Euclidean case, they are: (O, O); (O, I); (O, D); (I, I); (I, D) and (D, D). This classification was already obtained in \cite{Hacyan}, using a mixture of null tetrad and spinorial formalisms, and in \cite{Karlhede}, by means of splitting the four-dimensional Weyl tensor into two three-dimensional independent tensors of rank two and using the $SO(4|\mathbb{R})$ group to find canonical forms for these tensors.

\section{(2,2) Signature }\label{sec22}
In a real four-dimensional manifold of signature (2,2) it is possible to find a real frame $\{e_{1},e_{2},e_{3},e_{4}\}$ such that the only non-zero inner products between the basis vectors are $e_{1}^\mu e_{1\mu}=e_{2}^\mu e_{2\mu} = 1$ and  $e_{3}^\mu e_{3\mu}=e_{4}^\mu e_{4\mu}=-1$. Because of this, in such signature it is possible to construct a null tetrad frame made only of real vectors: $l=\frac{1}{\sqrt{2}}(e_{1}+e_{3}) \, , \, n=\frac{1}{\sqrt{2}}(e_{1}-e_{3})\,,\, m_1=\frac{1}{\sqrt{2}}(e_{2}+e_{4}) \, , \,m_2=\frac{-1}{\sqrt{2}}(e_{2}-e_{4})\,$. From the reality of these vectors and the reality of Weyl tensor it follows that all the Weyl scalars, (\ref{WeylScalars}), are real in this case. These scalars form a total of ten real independent components of the Weyl tensor, as it should be in a real manifold of dimension four.

But, differently from the other two signatures, there are no relations connecting the different Weyl scalars, so that all classification types are possible in (2,2) case. There are thus 21 possible algebraic types for the Weyl tensor in a four-dimensional real manifold of zero signature, they are (O,O); (O,I); (O,D);(O,II); (O,III); (O,N); (I;I); (I;D); $\ldots$, just as in the complex case. Where, as commented before, type (X,Y) is equivalent to type (Y,X). The canonical forms of these types are given in the same way as described at the end of section \ref{secLorentz}\footnote{It is possible that the bivector basis where the operators $C^\pm$ take the canonical forms is formed by complex bivectors. For example, when $C^+$ is type I$^+$ the null tetrad frame where $\Psi_0^+=\Psi_4^+=0$ may be composed by complex vectors, in particular in such a frame the Weyl scalars assume, in general, non-real values.}.

 A classification in signature (2,2) directly related to this one was obtained before in \cite{Law2}, by means of spinors. The relation of the present classification for $C^+$ and the one defined in \cite{Law2} is the following: I$^+$$\rightarrow\{1111\}$, $\{1\overline{1}11\}$ or $\{1\overline{1}1\overline{1}\}$; II$^+$$\rightarrow\{211\}$ or $\{1\overline{1}2\}$; D$^+$$\rightarrow\{22\}$ or $\{2\overline{2}\}$; III$^+$$\rightarrow\{31\}$ and N$^+$$\rightarrow\{4\}$. There are more Weyl tensor types in \cite{Law2} because there the reality of principal spinors is used to refine the classification. An analogous refinement could be done in the bivector approach, but probably this refinement do not bring any useful geometric information.

\section{Boost Weight Method}\label{secboost}
The most famous and, up to now, fruitful tensor classification scheme for higher-dimensional Lorentzian manifolds is the so called CMPP classification \cite{CMPP}, in particular when used to classify the Weyl tensor in four dimensions it reduces to the Petrov classification. This classification is based on the so called boost weight decomposition. In reference \cite{ColeyPSEUD} S. Hervik and A. Coley extended the boost weight classification to pseudo-Riemannian manifolds of arbitrary dimension, but as depicted in \cite{ColeyPSEUD} this classification can not be applied to Euclidian nor complex manifolds. It will be argued in the present section that this classification can be easily extended to complex manifolds and the results on real manifolds, including Euclidean signature, can be extracted from the complex case by imposition of suitable reality conditions. In particular, the boost weight classification for the four-dimensional Weyl tensor will be shown to be equivalent to the classification obtained here by the bivector approach.

When the metric of a real manifold has signature $(p,q)$ the isometry group on each point of the manifold (gauge group of the null frame) is $SO(p,q|\mathbb{R})$. But if the manifold is complexified the isometry group is enhanced to $SO(p+q|\mathbb{C})$. In \cite{ColeyPSEUD} the analysis is restricted to the real manifolds with real isometry groups, so that the boost weight classification as treated there does not applies to the complex manifolds nor to the Euclidean signature, since in this last case real null directions are not allowed. But this treatment can be easily generalized to include complex and Euclidean manifolds. In complex manifolds there is nothing special to do, just apply directly the boost weight decomposition while keeping in mind that the isometry group is $SO(n|\mathbb{C})$, where $n$ is the complex dimension of the manifold. In the Euclidean case the tangent bundle must be complexified so that the isometry passes from $SO(n|\mathbb{R})$ to $SO(n|\mathbb{C})$, after this the boost weight classification is done and at the end the reality condition must be imposed. As an example it will be shown below the classification of the Weyl tensor in Euclidean four-dimensional manifolds.

In the previous sections it was shown that in four dimensions it is always possible to locally find a null tetrad frame $\{l,n,m_1,m_2\}$, although it may be necessary to complexify the tangent bundle. In order to agree with the notation of \cite{ColeyPSEUD} let us define the following null basis: $l^1=n, n^1=l, l^2=-m_2$ and $n^2=m_1$. When the component of a tensor, $\Upsilon_A$, transforms as $\Upsilon_A\mapsto\lambda^{-r}\tau^{-s}\Upsilon_A$ under the boost $l^1\rightarrow\lambda l^1, n^1\rightarrow\lambda^{-1} n^1, l^2\rightarrow\tau l^2$ and $n^2\rightarrow\tau^{-1} n^2$ this component is said to have boost weight [$r$,$s$]. Now looking to the definition of the Weyl scalars $\Psi^+_a$, (\ref{WeylScalars}), it is easily seen that they have the following boost weights:
\begin{center}
\textmd{Boost Weights of the Weyl Scalars}\\
\begin{tabular}{|c|c|c|c|c|}

  \hline
  %after \\: \hline or \cline{col1-col2} \cline{col3-col4} ...
\,$\Psi^+_0$ $\rightarrow$ [2,2]\, &\, $\Psi^+_1$ $\rightarrow$ [1,1]\, &\,$\Psi^+_2$ $\rightarrow$ [0,0] \,&\,$\Psi^+_3$ $\rightarrow$ [-1,-1] \,&\,$\Psi^+_4$ $\rightarrow$ [-2,-2] \,\\
  \hline
\end{tabular}
\end{center}

The important thing to be observed in the above table is that the components of $C^+$ are such that the boost weights relative to the null vectors $\{l^1,n^1\}$ are always equal to the boost weights relative to $\{l^2,n^2\}$. So in the language of \cite{CMPP,ColeyPSEUD} the only allowed types for $C^+$ are [O,O], [I,I], [D,D], [II,II], [III,III] and [N,N]. These are respectively what in the present article was called types O$^+$, I$^+$, D$^+$, II$^+$, III$^+$ and N$^+$. Obviously the operator $C^-$ can have only the same six types. Finally, to apply the boost weight classification to the whole Weyl tensor, the classifications of $C^+$ and $C^-$ should be composed. It is then clear that in the Lorentzian and (2,2) signatures the boost weight classification for the Weyl tensor furnishes the same types as the bivector approach presented here. Obviously it must be considered, as was shown in section \ref{secLorentz}, that in the Lorentzian case $C^+$ is the complex conjugate of $C^-$, so that the type of the first operator is the same of the second one. Analogously the boost weight classification for the Weyl tensor for the complex and Euclidean cases, in the form explained above, follows easily and produces the same types as the bivector approach presented here. As an example let us treat explicitly the Euclidean signature.

As explained in the preceding paragraph, the operator $C^+$ can have only six types of boost weight: [O,O], [I,I], ... [N,N]. Analogously the operator $C^-$ can have only the same six types. In the particular case of the Euclidean signature, as shown in section \ref{secEuclid}, the Weyl scalars are such that  $\overline{\Psi^{\pm}_0} = \Psi^{\pm}_4$ and $\overline{\Psi^{\pm}_1} = -\Psi^{\pm}_3$. This implies that for both operators $C^+$ and $C^-$ the type [II,II] collapses to type [D,D] while types [III,III] and [N,N] collapse to type [O,O]\footnote{For example, if the operator $C^+$ is type [II,II] then, by definition, there exists a basis in which $\Psi^{+}_0=\Psi^{+}_1=0$. Now because of the identities $\overline{\Psi^{+}_0} = \Psi^{+}_4$ and $\overline{\Psi^{+}_1} = -\Psi^{+}_3$ that are valid in the Euclidean signature it follows that in this basis $\Psi^{+}_3$ and $\Psi^{+}_4$ must also vanish. So that the type of $C^+$ is indeed [D,D] instead of [II,II].}. So that in this signature both operators $C^+$ and $C^-$ must have one of the following three types: [O,O], [I,I] or [D,D]. This agrees perfectly with the result obtained by means of the bivector approach.

 \section{Conclusion}
A local classification for the Weyl tensor, based on the map of bivectors into bivectors that this tensor provides, was developed in all four-dimensional manifolds endowed with a metric and a Levi-Civita connection. The classification was presented in an unified way, so that the classification on real spaces can be seen as particular cases of the complex classification, where different signatures correspond to different choices of reality condition on the complex manifold. In the complex case and in the real one of signature (2,2) there are 21 possible classifications. In the Lorentzian and Euclidean signatures special relations appear when the reality condition is imposed so that just 6 types of classifications can be realized. It was also discussed the canonical forms of the Weyl tensor in the various classification types. Finally it was proved that trivially extending the boost weight classification of \cite{ColeyPSEUD} to include complex and Euclidean manifolds we get a classification scheme for the Weyl tensor that is equivalent to the bivector approach presented here in all signatures as well in complex manifolds.

\section*{Acknowledgments}
I am indebted to Professor Bruno G. Carneiro da Cunha for the patience in the revision of the manuscript and for the valuable proposals. I also want to thank Fábio M. Novaes Santos for the \LaTeX \, lessons, to Marcello Ortaggio for suggesting many references and to the referee for important suggestions. This work had CNPq\footnote{Conselho Nacional de Desenvolvimento Científico e Tecnológico} financial support. The final publication is available at \textit{link.springer.com} .


\begin{thebibliography}{20}%esse número diz que não passrá de  16 refeências
\bibitem{typeD} W. Kinnersley, \textit{Type D vacuum Metrics}, Journal of Mathematical Physics \textbf{10} (1969), 1195
\bibitem{Goldberg-Sachs} J. Goldberg and R. Sachs, \textit{A theorem on Petrov types}, General Relativity and Gravitation \textbf{41} (2009), 433.  This is a republication of original 1962 paper.
\bibitem{Stephani} H. Stephani et. al., \textit{Exact solutions of Einstein's field equations}, Cambridge University Press (2009)
\bibitem{PlebanskiBook} J. Pleba\'{n}ski and A. Krasinski, \textit{An introduction to general relativity and cosmology}, Cambridge University Press (2006)

\bibitem{Petrov} A. Z. Petrov, \textit{The classification of spaces definig gravitational fields}, General Relativity and Gravitation \textbf{32} (2000), 1665. This is a translated republication of original 1954 paper.
\bibitem{Plebanski75} J. Pleba\'{n}ski, \textit{Some solutions of complex Einstein equations}, Journal of Mathematical Physics \textbf{16} (1975), 2395
\bibitem{HallBook} G. S. Hall, \textit{Lecture notes on symmetries and curvature structure in general relativity}, World Scientific (2004)


\bibitem{Hacyan} S. Hacyan, \textit{Gravitational instantons in H-spaces}, Physics Letters \textbf{75A} (1979), 23
\bibitem{Karlhede} A. Karlhede, \textit{Classification of Euclidean metrics}, Classical and Quantum Gravity \textbf{3} (1986),L1 (Letter to the Editor).
\bibitem{Petrov-livro} A. Z. Petrov, \textit{Einstein Spaces}, Pergamon Press (1969), pages 99-101.
\bibitem{Law1} P. R. Law, \textit{Neutral Einstein metrics in four dimensions}, Journal of Mathematical Physics \textbf{32} (1991), 3039
\bibitem{Nurwoski2} A. Gover, C. Hill and P. Nurowski, \textit{Sharp version of the Goldberg-Sachs theorem}, Annali di Matematica Pura ed Applicata \textbf{190} Number 2 (2011), 295. Availabe at arXiv:0911.3364.
\bibitem{Law2} P. R. Law, \textit{Classification of the Weyl curvature spinors of neutral metrics in four dimensions}, Journal of Geometry and Physics \textbf{56} (2006), 2093
\bibitem{art2} C. Batista, \textit{A generalization of the Goldberg-Sachs theorem and its consequences}. Availabe at arXiv:1205.4666

\bibitem{ColeyBiv} A. Coley and S. Hervik, \textit{Higher dimensional bivectors and classification of the Weyl operator}, Classical and Quantum Gravity \textbf{27} (2010), 015002. Availabe at arXiv:0909.1160
\bibitem{CMPP} A. Coley, R. Milson, V.Pravda and A. Pravdová, \textit{Classification of the Weyl tensor in higher dimensions}, Classical and Quantum Gravity \textbf{21} (2004), L35. Availabe at arXiv:gr-qc/0401008v2.
\bibitem{ColeyScInvariant} A. Coley, S. Hervik and N. Pelavas, \textit{Spacetimes characterized by their scalar curvature invariants}, Classical and Quantum Gravity \textbf{26} (2009), 025013. Availabe at arXiv:0901.0791

\bibitem{ColeyPSEUD} S. Hervik and A. Coley, \textit{On the algebraic classification of pseudo-Riemannian spaces}. Availabe at arXiv:1008.3021
\bibitem{Bel} L. Bel, \textit{Radiation states and the problem of energy in general relativity}, General Relativity and Gravitation \textbf{32} (2000), 2047.  This is a republication of original 1962 paper.
\bibitem{Penrose} R. Penrose and W. Rindler, \textit{Spinors and space-time} vol.\textbf{1} and\textbf{ 2}, Cambridge University Press (1984 and 1986)





\end{thebibliography}
\end{document}